\documentclass{aa}  
\usepackage{graphicx}
\usepackage{txfonts}
\usepackage[]{hyperref}
\newcommand{\hi}{H\,{\sc i}}
\newcommand{\HI}{H\,{\sc i}}
\newcommand{\kms}{~km\,s$^{-1}$}
\newcommand{\msun}{~$M_{\odot}$}

\begin{document}

   \title{Apertif view of the OH Megamaser IRAS~10597+5926: \\ OH 18 cm satellite lines in wide-area \hi\ surveys}


   \author{Kelley M.~Hess
          \inst{\ref{astron},\ref{kapteyn}}\thanks{hess@astron.nl},
          H.~Roberts           \inst{\ref{cuboulder}},
          H.~D\'enes           \inst{\ref{astron}},
          B.~Adebahr           \inst{\ref{airub}},
          J.~Darling           \inst{\ref{cuboulder}},
          E.~A.~K.~Adams       \inst{\ref{astron},\ref{kapteyn}},
          W.~J.~G.~de~Blok       \inst{\ref{astron},\ref{uct},\ref{kapteyn}},
          A.~Kutkin            \inst{\ref{astron},\ref{lebedev}},
          D.~M.~Lucero         \inst{\ref{virginiatech}},
          Raffaella~Morganti   \inst{\ref{astron},\ref{kapteyn}},
          V.~A.~Moss           \inst{\ref{csiro},\ref{sydney},\ref{astron}},
          T.~A.~Oosterloo      \inst{\ref{astron},\ref{kapteyn}},
          R.~Schulz            \inst{\ref{astron}},
          J.~M.~van~der~Hulst  \inst{\ref{kapteyn}},
          A.~H.~W.~M.~Coolen   \inst{\ref{astron}},
          S.~Damstra            \inst{\ref{astron}},
          M.~Ivashina	\inst{\ref{chalmers}},			
          G.~Marcel~Loose      \inst{\ref{astron}},
          Yogesh~Maan          \inst{\ref{astron},\ref{uva}},
          \'A.~Mika             \inst{\ref{astron}},
          H.~Mulder             \inst{\ref{astron}},
          M.~J.~Norden          \inst{\ref{astron}},
          L.~C.~Oostrum        \inst{\ref{astron},\ref{uva}},
          M.~Ruiter             \inst{\ref{astron}},
          Joeri~van~Leeuwen    \inst{\ref{astron},\ref{uva}},
          N.~J.~Vermaas         \inst{\ref{astron}},
          D.~Vohl             \inst{\ref{astron}},
          S.~J.~Wijnholds      \inst{\ref{astron}},
            \and
          J.~Ziemke            \inst{\ref{astron},\ref{rugcit}}
          }

   \institute{ASTRON, the Netherlands Institute for Radio Astronomy, Postbus 2, 7990 AA, Dwingeloo, The Netherlands\label{astron}
         \and
             Kapteyn Astronomical Institute, University of Groningen, P.O. Box 800, 9700 AV Groningen, The Netherlands\label{kapteyn}
               \and
Center for Astrophysics and Space Astronomy, Department of Astrophysical and Planetary Sciences, University of Colorado, 389 UCB, Boulder, CO 80309, USA\label{cuboulder}
  \and
Ruhr University Bochum, Faculty of Physics and Astronomy, Astronomical Institute (AIRUB), Universit\"atsstrasse 150, 44780 Bochum, Germany\label{airub}
  \and
  Dept.\ of Astronomy, Univ.\ of Cape Town, Private Bag X3, Rondebosch 7701, South Africa\label{uct}
  \and
Astro Space Center of Lebedev Physical Institute, Profsoyuznaya Str. 84/32, 117997 Moscow, Russia\label{lebedev}
  \and
Department of Physics, Virginia Polytechnic Institute and State University, 50 West Campus Drive, Blacksburg, VA 24061, USA\label{virginiatech}
  \and
CSIRO Astronomy and Space Science, Australia Telescope National Facility, PO Box 76, Epping NSW 1710, Australia\label{csiro}
  \and
Sydney Institute for Astronomy, School of Physics, University of Sydney, Sydney, New South Wales 2006, Australia\label{sydney}
  \and
  Dept.\ of Electrical Engineering, Chalmers University of Technology, Gothenburg, Sweden\label{chalmers}
  \and
Anton Pannekoek Institute, University of Amsterdam, Postbus 94249, 1090 GE Amsterdam, The Netherlands\label{uva}
  \and
Rijksuniversiteit Groningen Center for Information Technology, P.O. Box 11044, 9700 CA Groningen, the Netherlands\label{rugcit}
             }

   \date{}
   
   \titlerunning{Apertif view of an OH megamaser}
   \authorrunning{K.~M.~Hess et al.}

 
  \abstract{
    We present the serendipitous detection of the two main OH maser lines at 1667 and 1665 MHz associated with IRAS~10597+5926 at $z_{\odot}=0.19612$ in the untargeted Apertif Wide-area Extragalactic Survey (AWES), and the subsequent measurement of the OH 1612 MHz satellite line in the same source.  With a total OH luminosity of $\log(L/L_{\odot}) = 3.90\pm0.03$, IRAS~10597+5926 is the fourth brightest OH megamaser (OHM) known.  We measure a lower limit for the 1667/1612 ratio of $R_{1612}>45.9$ which is the highest limiting ratio measured for the 1612 MHz OH satellite line to date.  OH satellite line measurements provide a potentially valuable constraint by which to compare detailed models of OH maser pumping mechanisms.  Optical imaging shows the galaxy is likely a late-stage merger.  Based on published infrared and far ultraviolet fluxes, we find that the galaxy is an ultra luminous infrared galaxy (ULIRG) with $\log(L_{TIR}/L_{\odot}) = 12.24$, undergoing a star burst with an estimated star formation rate of $179\pm 40$~\msun\ yr$^{-1}$.  These host galaxy properties are consistent with the physical conditions responsible for very bright OHM emission.
    Finally, we provide an update on the predicted number of OH masers that may be found in AWES, and estimate the total number of OH masers that will be detected in each of the individual main and satellite OH 18 cm lines.
    }


   \maketitle
%
\section{Introduction}

Despite their first discovery nearly four decades ago \citep{Baan82},
extragalactic OH masers are still relatively rare objects.  The majority are a million times more luminous than Galactic OH masers, earning them the name megamasers. It was recognized early on that OH megamasers (OHMs) are well correlated with FIR luminosity which is a signpost for the maser pumping mechanism (e.g.~\citealt{Baan99}): OHMs are essentially always found in luminous or ultra-luminous infrared galaxies ([U]LIRG; \citealt{Zhang14}), and they are associated with galaxy mergers, intense nuclear star formation activity \citep{Henkel91,Skinner97}, and regions of dense gas \citep{Darling07}.  
The strong correlation of OHMs with galaxy mergers, and the redshift information that comes automatically from a detection, make them a valuable tracer of the merger history of the Universe, and thus a test of hierarchical galaxy evolution models (e.g.~\citealt{Lo05, McKean09}).

The majority of extragalactic OH masers have been found through the detection of the 1667 MHz line by targeting bright IRAS sources with single dish radio telescopes (e.g.~NRAO 300 FT, \citealt{Baan92a}; Parkes Observatory, \citealt{StaveleySmith92,Norris89}; Arecibo Observatory, \citealt{Darling02}, \citealt{Fernandez10}; Green Bank Telescope, \citealt{Willett12}). Only of order a dozen OH megamasers have been studied at greater than arcsecond resolution.  Those that have show that the maser emission is confined within a 100 pc region of the nucleus.  Up to 30\% of the emission may be diffuse within the 100 pc region, while the majority of the emission is found in compact regions $\sim10$ pc in size (\citealt{Lo05}, and sources therein). 

In fact, at radio frequencies, OH has two main lines at 1665 and 1667 MHz and two satellite lines at 1612 and 1720 MHz \citep{Radford64}.
Extragalactic OHMs exhibit a broad range of 1667/1665 hyperfine ratios (from $R_H < 1$ to $R_H > 10$) with an average close to $R_H \sim 5$ \citep{McBride13}.
The variation observed in this property has been successfully modeled as a combination of a clumpy maser medium of overlapping clouds \citep{Parra05}, and far-infrared line overlap with pumping primarily occurring through the 53 $\mu$m line \citep{Lockett08}.  The former explains how compact masers (with typically high $R_H$ values) appear embedded in a diffuse emission (with low $R_H$) in high resolution very long baseline interferometry (VLBI) observations \citep{Lonsdale02}. Together these models account for how warmer dust is correlated with higher infrared luminosities \citep{Willett11,Lockett08}.

However, very little is known about the ratio between main and satellite lines in OHMs.  In local thermodynamic equilibrium (LTE), the emission line ratios between are expected to be 1 : 5 : 9 : 1 for 1612 : 1665 : 1667 : 1720 MHz.  Measuring the ratio of satellite to main lines provides a test for complex maser pumping models which predict the 1667/1612 ($R_{1612}$) and 1667/1720 ($R_{1720}$) ratios based on variation in the optical depth of the 1667 MHz line and the assumption that all lines have the same excitation temperature (\citealt{Lockett08,McBride13}, see also \citealt{Willett11}).
There are only four extragalactic detections of 1612 MHz in emission which is a requirement to measure the 1667/1612 line ratio: Arp 220 \citep{Baan87}, Arp 299/IC 694 and Mrk 231 \citep{Baan92a}, and IRAS F15107+0724 \citep{McBride13}.  (A tentative detection of II Zw 96/IRAS 20550+1655 by \citealt{Baan89} was refuted by \citealt{McBride13}). 
Their corresponding measurements of the 1667/1612 ratio span $R_{1612}\sim7-29$ over a relatively narrow range in 1667 MHz optical depth \citep{Baan92a,McBride13}.
In addition, M82 is detected with a complex mix of 1612 MHz emission and absorption \citep{Seaquist97} and three extragalactic OH masing sources have been detected in 1612 MHz absorption: NGC~253 \citep{Frayer98}; PKS~1413+135 \citep{Darling04}; PMN~J0134-0931 \citep{Kanekar05}.  In all of the cases when 1612 MHz is detected in absorption, the 1720 MHz line is found in conjugate emission.  In contrast, while Arp~220 is detected in 1612 MHz emission, the 1720 MHz line is seen in absorption \citep{Baan87}. These complex cases have not yet been captured in the discussion of the maser pumping models.

In addition to probing the state of the masing gas, satellite lines of OH are used to measure the universality of fundamental constants at high redshift \citep{Varshalovich95,Darling03,Chengalur03}.  The four OH lines have different dependencies on the fine structure constant, $\alpha = e^2 / \hbar c$, the electron-proton mass ratio, $\mu = m_e/m_p$, and the proton gyromagnetic ratio, $g_p$ \citep{Kanekar05}.  Early quasar studies claimed a possible changing $\alpha$ (e.g.~\citealt{Webb01}).  While no evidence for changing fundamental constants has been found to date from cm lines, only two OH sources are known which have been used for such studies (PKS 1413+135, $z = 0.24671$; PMN~J0134-0931, $z=0.765$; \citealt{Darling04,Kanekar12,Kanekar18}), and a small number of sources using other lines (e.g.~ammonia, \citealt{Flambaum07}; or methanol, \citealt{Jansen11}). 

Large volume radio spectral line surveys at 1.4 GHz, typically aimed at targeting the 21 cm line of neutral hydrogen, \HI, provide an unprecedented opportunity to discover and study the rare, but cosmologically important extragalactic OH megamasers (e.g. \citealt{Briggs98, Suess2016}).  The large bandwidth of new and upgraded instruments undertaking these surveys will allow the simultaneous measurement of main and satellite lines, while also detecting or putting limits on the \hi\ emission and absorption strength \citep{Gupta20}.  Apertif is one of these upgraded survey machines: a new phased array feed (PAF) instrument on the Westerbork Synthesis Radio Telescope undertaking commensurate spectral line, continuum, and polarization surveys in the frequency range 1130-1430 MHz (\citealt{Adams19}; van Capellen et al., in prep; Hess et al., in prep).  In addition to directly detecting \HI\ in galaxies out to $z\sim0.07$ (and through \hi\ stacking or absorption out to the bandwidth limit, $z\sim0.2$), the observations are capable of detecting extragalactic OH masers where the 1667 MHz line has been redshifted to $0.166<z<0.476$. 

In this paper, we present the untargeted, serendipitous detection of IRAS~10597+5926 in both the 1667 and 1665 main lines, and a measurement of the 1612 MHz satellite line with a integrated signal-to-noise ratio of 2.2$\sigma$ which we treat as an upper limit.  
We use published infrared (IR) and far ultraviolet (FUV) fluxes available in the NASA/IPAC  Extragalactic Database\footnote{http://ned.ipac.caltech.edu; The NASA/IPAC Extragalactic Database (NED) is funded by the National Aeronautics and Space Administration and operated by the California Institute of Technology.} to calculate the host galaxy properties including star formation rate and stellar mass, and confirm that IRAS~10597+5926 is an ultra luminous infared galaxy (ULIRG).
The 1612 MHz emission measurement, is the highest 1667/1612 ratio lower limit measured to date, and the most stringent constraint on existing maser pumping models.
This measurement demonstrates the power of upcoming untargeted, large-volume \HI\ surveys, where the broad bandwidth is valuable for simultaneously observing multiple OH lines either directly or through the potential of OH stacking.  The detection of IRAS~10597+5926 also showcases the automated source finding/inspection tools under development which make discoveries within these surveys, and the harvesting of data from spectral line sources possible.

Throughout this paper we assume a $\Lambda$CDM cosmology with $H_0=70$ km s$^{-1}$ Mpc$^{-1}$, $\Omega_M=0.27$, $\Omega_{\Lambda}=0.73$.

\begin{figure*}
    \centering
    \includegraphics[width=\textwidth]{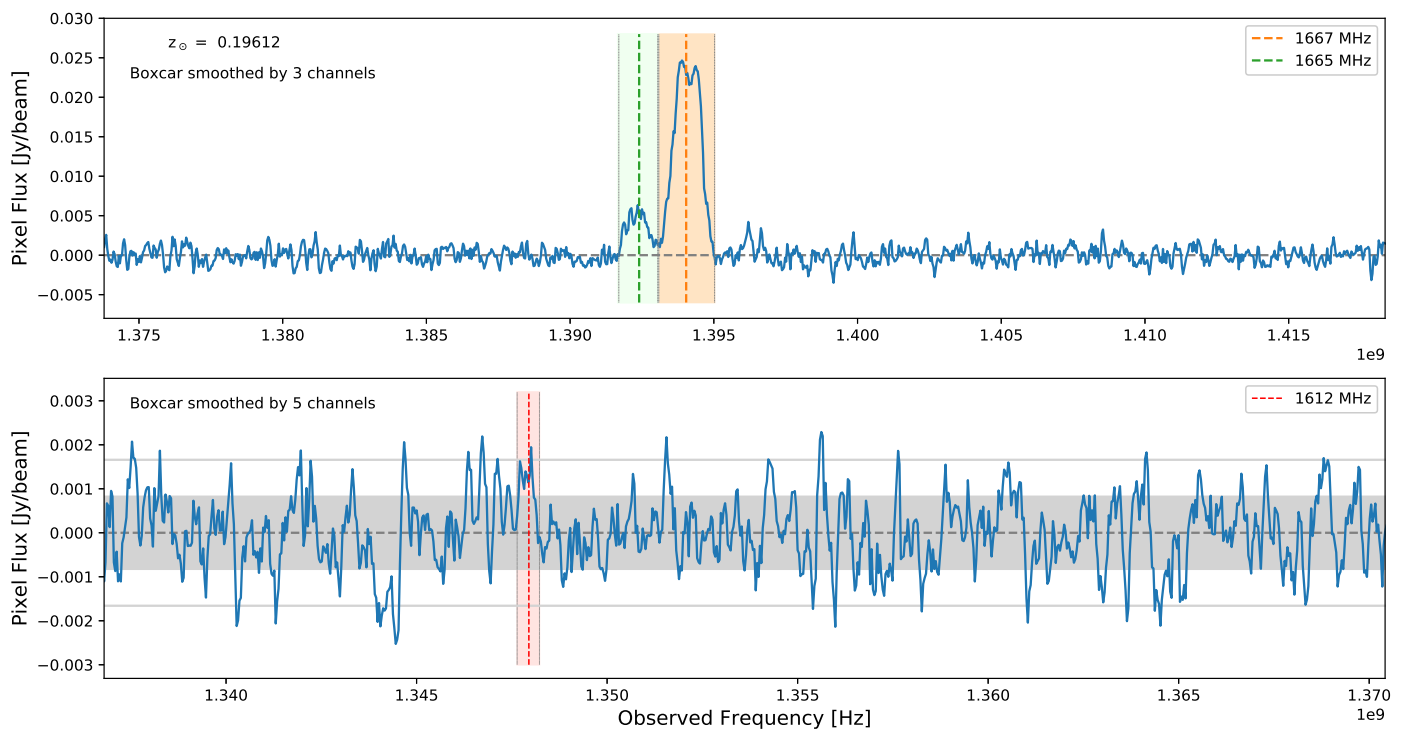} \caption{Top: spectrum of the 1665 and 1667 MHz main lines.  Bottom: spectrum of the 1612 MHz satellite line.  The 1667 MHz line center is provided by a fit to the line profile by SoFiA.  The 1665 and 1612 MHz lines are plotted at the expected frequency based on the 1667 MHz line redshift. The orange/green/red shaded regions indicate the channel range used to calculate the intergrated flux of each line which correspond to widths of 433, 308, 138\kms, respectively.   Grey shaded region in the bottom plot shows the $\pm1$ sigma noise range of the smoothed data; Grey line indicates $\pm2$ sigma.}
    \label{fig:spectra}
\end{figure*}

\section{Apertif data}

\subsection{Data reduction, imaging, and source finding}
The area of sky containing IRAS~10597+5926 was observed in October 2019 (ObsId 191003042) as part of the ongoing Apertif wide-area extragalactic imaging survey, AWES.  The Apertif spectral line, continuum, and polarization surveys will be described in detail in Hess et al (in prep).  In short, Apertif is a phased array feed instrument upgrade to the Westerbork Synthesis Radio Telescope, which increases the field-of-view to 6.4-8 deg$^2$ and the spectral bandwidth to 300 MHz from 1130-1430 MHz.  Every AWES survey pointing is observed for 11.5 hours.  The data are reduced through an automated data reduction pipeline called \textit{Apercal} (\citealt{Schulz20}, Adebahr et al, in prep) which produces cleaned multi-frequency continuum (Stokes I) images; polarization (Stokes V) images and (Stokes Q- and U-) cubes; and continuum-subtracted spectral line dirty image cubes with 12.2--36.6 kHz resolution.  The raw, calibrated, and imaged data from the first 12 months of observing (1 July 2019 - 30 June 2020) are publicly available as part of the first Apertif data release\footnote{PID: \url{http://hdl.handle.net/21.12136/B014022C-978B-40F6-96C6-1A3B1F4A3DB0}} (Apertif DR1; Adams et al, in prep).

As part of ongoing Apertif work towards developing a catalog of \hi\ spectral line sources, the data was run through a spectral line source finding pipeline.  The pipeline is still under development and will be described in a future paper.  In short, we use a well-tested software application, SoFiA\footnote{\url{https://github.com/SoFiA-Admin/SoFiA}} \citep{Serra15} to find candidates spectral line sources.  SoFiA finds sources by searching for emission at multiple angular and velocity resolutions after smoothing the data cube with 3D kernels specified by the user. At each resolution, voxels are detected if their absolute value is above a threshold given by the user (in units of the rms noise). The final 3D mask is the union of the 3D masks constructed at the various resolutions. Objects are cleaned to 0.5$\sigma$ within masks defined by SoFiA encompassing the source.  The data cubes are compound beam-corrected using standard tasks in the Miriad software package \citep{Sault95,Sault11}, and adjusted to the NVSS flux scale (see documentation in the Apertif data release).  Final data products which include moment maps and spectrum plots are created using Astropy v2.0 \citep{astropy18}.  The final data products are then inspected by-eye and unusual objects are flagged for potential science interest.

In the context of OH masers, the 300 MHz frequency range covered by the Apertif surveys provide a window to observe the 1667 MHz line between redshifts of $0.16599<z<0.47554$.  This redshifted OH maser line can effectively masquerade as an \HI\ source in the nearby Universe with similar flux densities and line widths (e.g.~\citealt{Briggs98,Suess2016,Haynes18}).  A spectral line source finding pipeline works effectively for both \hi\ and OH lines because it cannot distinguish between the two.  In this way, we serendipitously detected the 1667 MHz line of IRAS~10597+5926 (J2000 11:02:46.9 +59:10:37), redshifted to $z_{\odot}=0.19612$, while inspecting candidate \HI\ detections.
In this case the OH nature of the source was clear because the 1665 MHz line was also clearly detected.  A literature search found that the 1667 MHz line was presented in an unrefereed IAU Conference Proceedings by \citet{Willett12}, but the 1665 MHz line was not discussed and may have been left unreported due to a poorly determined or perhaps over-subtracted baseline.

We note that unfortunately frequencies below $\sim1300$ MHz, corresponding to the lower half of the Apertif band, are strongly impacted by radio frequency interference from satellites.  As a result, Apertif data to date are only processed between 1430-1280 MHz, decreasing the effective available redshift range to $0.16599<z<0.30262$ for OH line detections.  Table \ref{tab:redshifts} lists the redshift range in which each of the OH maser lines are visible to Apertif.

\subsection{Post-processing}

In the field of IRAS~10597+5926 are two bright continuum sources at roughly $770\pm15$ mJy (20 arcseconds away) and $580\pm10$ mJy (30 arcseconds away) 
which are not perfectly modeled by the calibration in the \textit{Apercal} pipeline, leaving faint concentric rings around the continuum sources.   These rings vary slowly in frequency and are faintly evident even in the line cubes.  To remove the artefacts, we mask the OH maser source and re-fit the baseline at every spatial pixel in the cubes with a 5-knot natural cubic spline\footnote{\url{https://github.com/madrury/basis-expansions/}} along the frequency axis.  The line cubes consist of 1216 channels of approximately 7 km s$^{-1}$, so a spline segment spans more than 1700 km s$^{-1}$, ensuring that every spline segment is dominated by emission free channels.  The resulting data cube has a noise of 1.27 mJy beam$^{-1}$.

The source finding pipeline was then re-run on the new data cube, which identified the 1667 MHz and 1665 MHz emission as separate line sources.  For the subsequent analysis we extracted a pixel spectrum at the location of the 1667 MHz peak in the total intensity map.  The spectrum was then boxcar smoothed by 3 channels (top panel of Figure \ref{fig:spectra}).  For the redshift of the system, we take the line center calculated by SoFiA for the 1667 MHz line and use it to calculate the expected frequencies for the 1665 MHz and 1612 MHz lines.  To calculate the line flux we integrate the pixel spectrum over $\pm27$ channels (433\kms) of the 1667 MHz line center, and $\pm$19 channels (308\kms) of the 1665 MHz line which correspond to the widths of the SoFiA masks for each source.

\subsection{1612 MHz data}

Based on the systemic redshift measured by SoFiA, we calculated that the 1612 MHz satellite line of IRAS~10597+5926 would also fall in the Apertif bandwidth, redshifted to 1347.9469 
MHz.  We applied the same spline fitting technique to this data cube to remove residual artefacts.  The resulting cube has a noise of 1.30 mJy beam$^{-1}$. We extracted the pixel spectrum at the position of the peak of the 1667 MHz total intensity map.  The spectrum was then boxcar smoothed by 5 channels (bottom panel of Figure \ref{fig:spectra}). 
To calculate the line flux we integrate over $\pm8$ channels (138\kms) of the expected 1612 MHz line center.  This corresponds, subjectively, to a symmetric extent roughly the width of the brightest part of the 1665 and 1667 MHz lines, and where we would expect the 1612 MHz line to be brightest if it is present (see also the gray shaded region in Figure \ref{fig:matchedfilter}).

We have done a number of tests to understand the significance of a measurement at the exact location of the predicted 1612 MHz line, including correlating the Apertif spectrum with different matched filters, and conducting a Bayesian analysis with a strong redshift prior.  These tests are described in Appendix \ref{sect:app}.  The matched filter tests modestly strengthen the case for a detection of the 1612 MHz line.  However, the Bayesian analysis suggests that there is roughly a 1 in 3 chance of a detection at that location given the noise of the spectrum.  As a result, we describe our measurement of the integrated flux of the 1612 MHz line as an upper limit.

\begin{table}[]
    \caption{Range of redshifts for OH maser lines visible to Apertif}
    \centering
    \begin{tabular}{c c c c c}
         \hline\hline
         OH Line & 1430 MHz & 1280 MHz & 1130 MHz \\
         \hline
         1612 MHz & 0.12743 & 0.25956 & 0.42675 \\
         1665 MHz & 0.16462 & 0.30110 & 0.47381 \\
         1667 MHz & 0.16599 & 0.30262 & 0.47554 \\
         1720 MHz & 0.20316 & 0.34416 & 0.52259 \\
         \hline
    \end{tabular}
    \tablefoot{Apertif observes frequencies up to 1430 MHz which defines the low redshift limit of the Apertif bandwidth; 1280 MHz is the redshift upper limit for frequencies relatively unimpaired by interference; 1130 MHz is the low frequency (high redshift) limit for the Apertif bandwidth.}
    \label{tab:redshifts}
\end{table}

\section{Results}
\label{sect:results}

We measure all OH maser emission lines of the $^2\Pi_{3/2}(J=3/2)$ state that are within in the Apertif bandwidth: the 1612, 1665, and 1667 MHz lines, redshifted to $z_{\odot}=0.19612$.  The two main lines are clearly visible in the top panel Figure \ref{fig:spectra}.  They have an integrated signal-to-noise ratios (SNR) of 12 and 57, respectively.  The bottom panel of Figure \ref{fig:spectra} shows the spectrum and predicted location of the 1612 MHz satellite line.  The peak of the spectrum within the estimated profile width is measured to be 2.3 sigma above the noise, while the integrated line profile over the same width has an SNR of 2.2.  The calculations of the integrated SNRs takes into account the correlated noise due to smoothing the spectra.  Only the 1720 MHz satellite line remains outside the frequency range of Apertif at 1438 MHz.  

Given our chosen cosmology, IRAS~10597+5926 sits at a luminosity distance of $D_L=963$ Mpc.  We assume isotropic emission and measure the luminosity in the 1667 and 1665 MHz lines to be $\log(L/L_{\odot}) = 3.83\pm0.01$ and $3.09\pm0.04$, respectively, and a 1612 MHz upper limit of $\log(L/L_{\odot}) \le 2.2\pm0.2$. Our errors reflect the uncertainty in the flux measurements based on the noise in the spectra.   The total luminosity in the 1665 and 1667 lines, $\log(L/L_{\odot}) = 3.90\pm0.03$, makes IRAS 10597+5926 the 4th most luminous OH maser source known, and places it on the margin of classification as an OH gigamaser.  Our flux seems to be in good agreement with that of \citet{Willett12} who measure $\log(L/L_{\odot}) = 3.85$ for the 1667 MHz line using the GBT, although the cosmological parameters used to calculate their value are not published.

\begin{figure}
    \centering
    \includegraphics[width=0.48\textwidth]{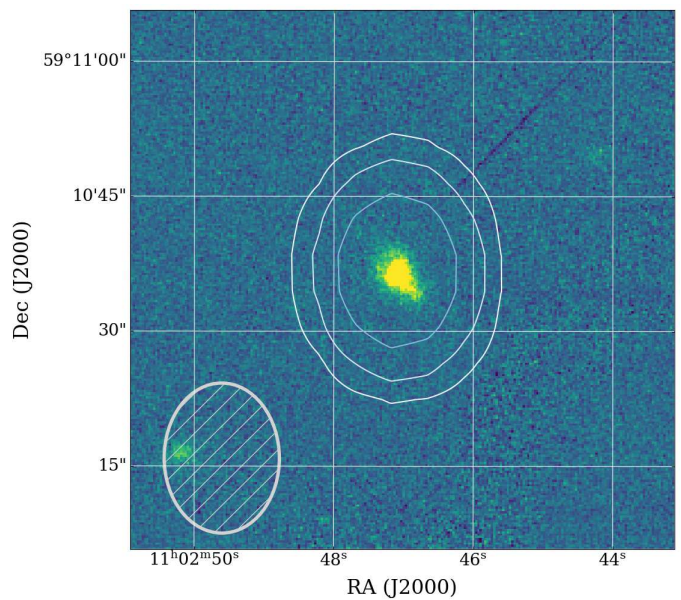}
    \caption{Continuum contours 6, 12, 24$\times10^{-4}$~Jy beam$^{-1}$ on a PanStarrs g-band image.  The radio source is unresolved by the \textbf{$13\times 17$} arcsec beam of Apertif (hatched grey ellipse in the bottom left corner).  The continuum image has a noise of 33 $\mu$Jy beam$^{-1}$. The disturbed optical appearance is consistent with the galaxy in the late stages of a merger.}
    \label{fig:maps}
\end{figure}

Figure \ref{fig:maps} shows an optical PanSTARRS\footnote{\url{https://panstarrs.stsci.edu/}} \citep{Chambers2016} g-band image with the Apertif 1.36 GHz radio continuum contours.  The radio continuum image has an rms of 33 $\mu$Jy beam$^{-1}$ at the location of the IRAS~10597+5926 and the continuum source associated with the maser is a $6\pm1$ mJy point source at the resolution of Apertif, $13 \times 17$ arcsec.  The ground-based images show a disturbed optical morphology which we interpret as an advanced merger as is seen in a majority of ULIRG sources for which there is sufficient imaging (e.g.~\citealt{Clements96,Farrah01}).

Among powerful OHMs, IRAS~10597+5926 is unique in that the 1667 and 1665 MHz lines are relatively narrow and distinct, whereas the brightest gigamasers show blended emission features for the 1667 and 1665 MHz lines spanning over 1000\kms\ \citep{StaveleySmith89,Baan92b,Darling01}. The brightest known IR 14070+0525 spans 2400~\kms. \citet{Baan92b} suggests this is either the result of a violent merger, an AGN outflow, or a 500~\kms\ rotating disk around a supermassive blackhole in which the 1667 and 1665 lines are blended \citep{Baan01}.  We note that the 1667 MHz line appears to have two distinct peaks, however VLBI imaging is required to say anything about the configuration of the masing material, or narrowness of the OH lines in IRAS~10597+5926.

\section{Discussion}

This first serendipitous OHM detected from the AWES survey of Apertif demonstrates that future wide-area \HI\ surveys will not only be a useful way to discover new extragalactic OH masers \citep{Briggs98}, but with their broad spectral bandwidth, provide simultaneous detections or constraints on OH satellite lines. Below we present the host galaxy properties, we discuss the properties of the OHM in the context of what is known about satellite lines, and we present a prediction for the discovery space which will be revealed with AWES.

\subsection{Host Galaxy Properties}

The infrared (IR) luminosity and star formation rate (SFR) of the host galaxy are tied to the physical conditions thought to be responsible for OHM activity. IRAS~10597+5926 is a very luminous OHM, so one would expect it to be very luminous in the mid- and far-IR.  We use publicly available archival data to confirm the extreme IR and SFR properties of this system.

IRAS~10597+5926 is detected in FUV by GALEX \citep{Buat07}, in all four WISE infrared bands images at 3.4, 4.6, 12, 22 $\mu$m \citep{Zhang14}, and in IRAS 60 and 100 $\mu$m band images \citep{Moshir90}.  For the WISE bands we use $k$-corrected fluxes based on a galaxy template and SED fitting \citep{Jarrett19}. For IRAS bands this is not available so for them and the GALEX data we take into account a zeroth order $k$-correction of $1/(1+z)$ dependence with redshift and represent the additional uncertainty in the error. 

We estimate the star formation rate (SFR) using two methods from independent subsets of the available IR and UV data.  First, we calculate the total IR (TIR) flux from IRAS as outlined in \citet{Sanders96}.  IRAS 12 and 25 $\mu$m fluxes are upper limits.  Ignoring these values we find a lower limit to be $\log(L_{TIR}/L_{\odot}) > 12.24$.  By including them we find an upper limit to the total IR luminosity of $\log(L_{TIR}/L_{\odot}) < 12.37$. In fact, the WISE W3 and W4 fluxes suggest the 12 and 25 $\mu$m contribution may be an order of magntiude lower than that suggested by the above mentioned upper limit.  To calculate the SFR, we combine Equations 11 and 12 of \citet{Kennicutt12}\footnote{See also \citealt{Murphy11,Hao11} for the constants used in these equations.} to derive a FUV + IR SFR from the $L_{TIR}$ lower limit of $179\pm40$~\msun\ yr$^{-1}$.  Alternatively, using the $k$-corrected WISE 22 $\mu$m band (W4) and calibration from \citet{Cluver17}, we find an SFR of $108\pm42$~\msun\ yr$^{-1}$.  We note that W4 is a better estimate of SFR than W3 in ULIRGs because of, among other things, strong Si dust absorption in the W3 band (T.~Jarrett, private communication).
Finally, we estimate a galaxy stellar mass based on the WISE 3.4 $\mu$m (W1) and the mass-to-light ratio from \citet{Cluver14} to be $\log(M_{*}/M_{\odot}) = 10.65\pm0.14$.  

The lower limit to the total IR luminosity, which is calculated exclusively from the 60 and 100 $\mu$m far IR bands, firmly qualifies IRAS~10597+5926 as a ULIRG \citep{Houck85}. Although we calculate a range of star formation rates based on far IR+FUV or mid-IR, both estimates put the rate at well over 100\msun\ yr$^{-1}$, and indeed the WISE 3.4-4.6 $\mu$m (W1-W2 = $0.987\pm0.053$) and 4.6-12 $\mu$m (W2-W3 = $4.315\pm0.058$) colours place it firmly in the range of starbursting galaxies \citep{Cluver14}. This is consistent with the physical conditions responsible for OHM activity.

\subsection{OH line ratios} \label{sect:lineratios}
In local thermodynamic equilibrium (LTE), the OH lines are expected to have ratios of $1612:1665:1667:1720 = 1:5:9:1$ \citep{Radford64,Robinson67,Henkel91}.  
However a broad range of values has been measured in megamasers with the highest resolution images revealing that compact emission tends to have 1667/1665 emission ratios closer to $R_H=5$ \citep{Lonsdale02}.

\begin{figure}
    \centering
    \includegraphics[width=0.48\textwidth]{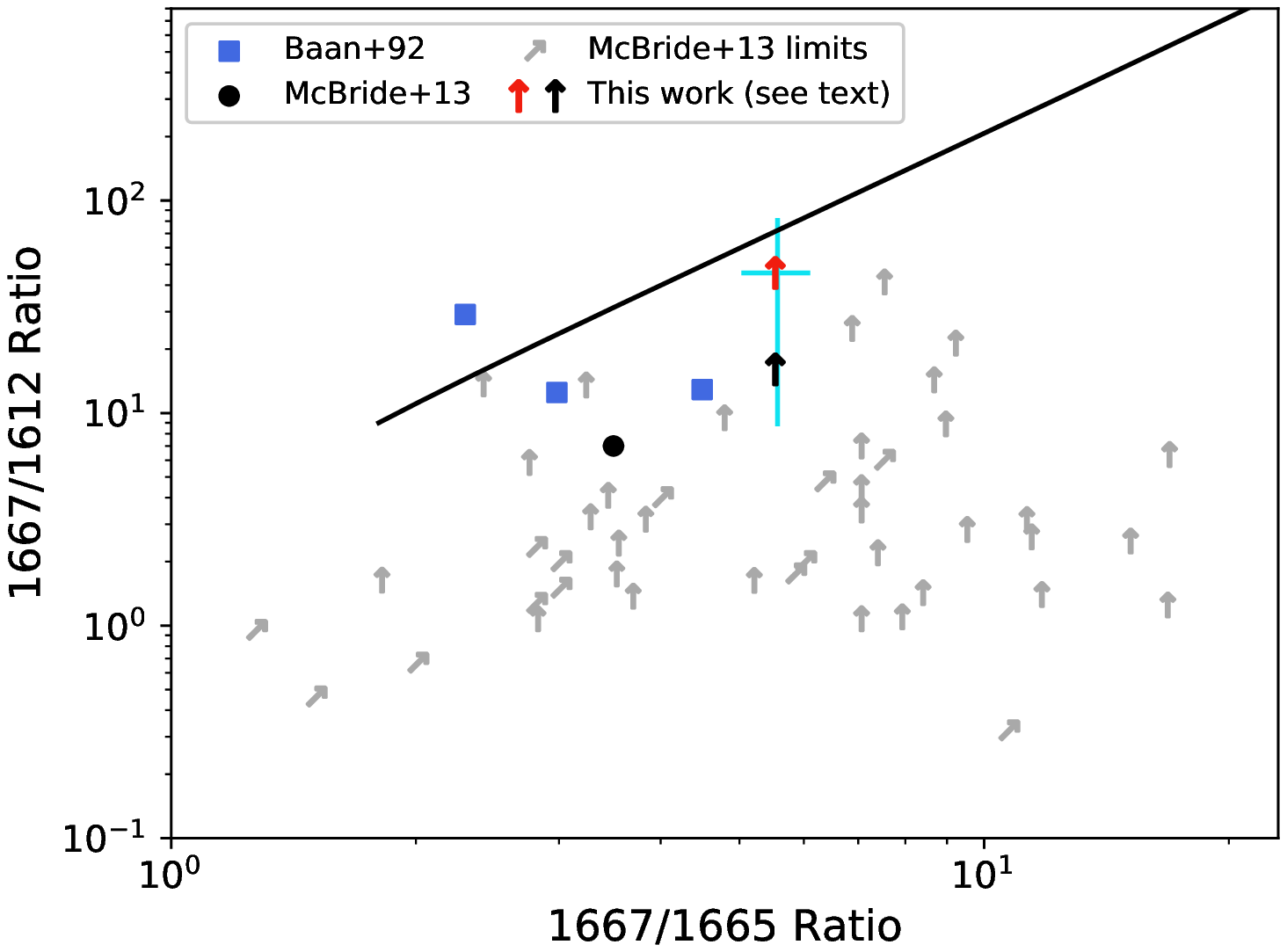}
    \caption{The OH color-color plot for the 1612 MHz line, adapted from \citet{McBride13}.  Blue squares come from the literature, collated by \citet{Baan92a} in their Table 2(a); black circle and gray arrows are the detection and upper limits from \citet{McBride13}, respectively. The red arrow is the $R_H$ ratio and $R_{1612}$ upper limit from this work; cyan cross shows our uncertainties; the black arrow is the upper limit calculated using the \citet{McBride13} methodology (see Section \ref{sect:lineratios}). The black line is the predicted ratios assuming equal excitation temperatures for all OH lines, with varying 1667 MHz optical depth \citep{Lockett08, McBride13}.}
    \label{fig:mcbride}
\end{figure}

For IRAS~10597+5926, we find (limiting) line ratios of $1:8.3:45.9$.  This corresponds to a hyperfine ratio between 1667/1665 of $R_H=5.5\pm0.5$.   The 1667/1612 ratio, $R_{1612} > 45.9$, is the largest line ratio limit ever measured for this satellite line.  Figure \ref{fig:mcbride}, adapted from \cite{McBride13}, shows our upper limit (red arrow) versus all existing measurements and limits in the literature\footnote{We note that the original \cite{McBride13} plot had two black points despite their table only listing one detection.  Ultimately a wiggle in the spectrum of IRAS 11028+3130 was deemed unreliable and updated in the table as an upper limit, but accidentally left in the plot (J.~McBride, private communication).  We have corrected it here.}.  For direct comparison, we also calculate an upper limit (black arrow) based on the methodology of \cite{McBride13}, who use $F_{1612}= \sigma_{1612} \, \Delta\nu_{1667}$ where $\sigma_{1612}$ is the rms of the 1612 MHz spectrum and $\Delta\nu_{1667}$ is the width of the 1667 MHz line containing 75\% of the line emission.  For IRAS 10597+5926, $\Delta\nu_{1667}$ is $\pm13$ channels--more than 60\% wider than our estimate. The black arrow falls within the uncertainty of our estimate, but we note that this method is (1) independent of the properties of spectrum at the predicted location of the line and (2) independent of the shape of the 1667 MHz line, which may suggest a more individualized approach.  In fact, if we integrate the 1612 MHz spectrum over $\pm13$ channels, we get the same lower limit as indicated by the red arrow.

The solid black line in Figure \ref{fig:mcbride} shows the predicted line ratios in ``color-color'' space if all OH lines in the $^2\Pi_{3/2}(J=3/2)$ state have equal excitation temperatures \citep{Lockett08,McBride13}.  If the $R_{1612}$ ratio deviates from this line, it may indicate secondary maser pumping mechanisms are important and should be explored in future models.  On face value, our measurement is among the strongest constraints that OHMs may still be globally consistent with the equal excitation temperature hypothesis, although a deviation from this is also within the uncertainties of the measurement.

\subsection{Updated predictions for Apertif, and a word on 1612 co-detections and stacking.}

The Apertif surveys have recently been guaranteed observing time through December 2021, for a total of 30 months of survey operations.  
Previous papers have estimated the Apertif OHM 1667 MHz detection rate based on specifications before the surveys were operational (e.g.~\citealt{McKean09}).  Here, we provide an update on the expected number of detections for AWES based on Roberts, Darling, \& Baker (2021; submitted). 

The estimate combines the OH luminosity function and assumed merger evolution rate, with the redshift range, area, and sensitivity of the radio survey.  The average OH line has a width of about 150 km s$^{-1}$ which covers roughly $\sim$23 Apertif channels.  We assume the data quality is consistent with the first data release which achieves an average noise of approximately 1.3 mJy beam$^{-1}$ in the \hi\ cubes at 36.6 kHz spectral resolution, and we require the OH megamasers to be 5$\sigma$ detections. Though there is much speculation of the exact value, we assume a galaxy merger evolution rate ($\gamma$) of 2.2 (\citealt{Lotz2011}). This value is associated with major mergers which is the source of most OHMs.

In addition, an extrapolation of the calculation for the 1667 MHz detections can be done to determine the number of satellite lines that may be detected by Apertif. Assuming all OHMs produce every satellite line and the median line ratios from \cite{McBride13}, we can say 1667/1665 line ratio is about 5 and that the 1667/1612 and 1667/1720 ratios are both around 20. This means we need 5 times more sensitivity for the 1665 MHz line and 20 times more sensitivity to detect 1612 and 1720 MHz lines. Using this and the adjusted redshift ranges (see Table \ref{tab:redshifts}), we can predict the number of detections for each line.

The exact survey plan for 2021 is under discussion, so we present the numbers for the survey observing to through the end of 2020, and two possible strategies for 2021 (Table \ref{tab:NOHs}).  Column (1) is the frequency of the OH line in MHz. Column (2) is the assumed line ratio with respect to the 1667 MHz line and is the median line ratio from \citet{McBride13}.  Column (3) is the number of OHMs expected from AWES observations in the first 18 months of the survey through the end of 2020 covering about 1500 deg$^2$.  Column (4) is the total number of OHMs assuming that AWES continues to build up sky coverage in the same way as it has for the previous 18 months for an estimated coverage of 2500 deg$^2$. Column (5) assumes that AWES re-observes existing coverage of approximately 1500 deg$^2$ to get 1.4 times more sensitivity in spectral line maps.  In both future cases, we also estimate the number of OHMs that would be detected between 1430-1530 MHz in 1000-1500 deg$^2$ assuming the Apertif frequency coverage will shift to avoid RFI beyond 1280 MHz (see caption). 

We estimate the total number of Apertif 1667 MHz OH megamaser detections to be of order 110-150 (depending on the 2021 strategy and including a broader freqeuency range in that 12 months), increasing the number of known OHMs by over 100\%. 
The large error bars are due to the limited number of known OHMs so as more are discovered, these predictions can be tightened.

This prediction is much higher than the approximately 19 OHMs that have been reported by the Arecibo Legacy Fast ALFA Survey (ALFALFA; \citealt{Haynes18}) which covered $\sim 7000$ deg$^2$. However, it is very likely that ALFALFA has yet to discover all potential OHMs and \cite{Suess2016} present 60 ambiguous sources detected by ALFALFA's 40\% data release alone. Further, Apertif's sensitivity is about a factor of two improved over ALFALFA (1.86 mJy beam$^{-1}$ over 48.8 kHz after Hanning smoothing; \citealt{Haynes18}): an important consideration for detecting OHMs since they must be redshifted, and therefore fainter, to be at the proper frequency for detection. 

\begin{table}[]
    \caption{OH line detection predictions for AWES to 2020 and beyond.}
    \centering
    \begin{tabular}{c c c c c}
         \hline\hline
         OH Line &  Line     & AWES$^{b}$ &  AWES$^{b,c}$ & 1.4x AWES$^{b,c}$ \\
         MHz   & Ratio$^{a}$ & 1500 deg$^2$ & 2500 deg$^2$ & 1500 deg$^2$ \\
         (1)     & (2)         & (3)          & (4)          & (5)          \\
         \hline
         1612  & 20 & \phantom{0}2$^{+1}_{-1}$     & \phantom{00}4$^{+1}_{-1}$     & \phantom{00}5$^{+1}_{-1}$ \\ [0.1cm]
         1665  & 5  &           17$^{+4}_{-3}$     & \phantom{0}28$^{+6}_{-5}$     & \phantom{0}25$^{+5}_{-4}$ \\ [0.1cm]
         1667  & 1  &           76$^{+17}_{-13}$   &           127$^{+27}_{-21}$   & 103$^{+24}_{-19}$           \\ [0.1cm]
         1720  & 20 &          0.5$^{+0.2}_{-0.2}$ & \phantom{0}0.9$^{+0.4}_{-0.3}$ & \phantom{00}2$^{+1}_{-1}$ \\ [0.1cm]
         \hline
    \end{tabular}
    \tablefoot{$^{a}$The assumed line ratio is in relation to the primary 1667 MHz line and is the median line ratio from \cite{McBride13}.  $^{b}$The number of detections are calculated for the frequency range 1280-1430 MHz.  $^{c}$For AWES projections beyond 2020 which may shift the frequency coverage (see text), we predict Apertif would detect an additional $\sim$12-20 (4-6) OHMs in the 1667 (1665) MHz line and an additional $\sim$1-2 OHMs in each of the satellite lines between 1430-1530 MHz, on top of what is listed above.}
    \label{tab:NOHs}
\end{table}

For undetected 1612 MHz lines, one may attempt to stack OHMs in bins of $R_H$, redshift, or other physically meaningful parameters.  Given the expected number of 1667 MHz detections, and assuming the noise goes down as the square root of the number of objects, we expect stacking would modestly improve the detection threshold by a factor of a few in each bin.  

\section{Conclusions}

We present the Apertif Wide-area Extragalactic Survey OH megamaser detection of IRAS~10597+5926.  The main lines at 1667 and 1665 MHz were found redshifted to $z_{\odot}=0.19612$ by an automated spectral line source finding pipeline.  The optical image suggests IRAS~10597+5926 is a late-stage merger.  From infrared imaging, we show that the host galaxy is a ULIRG, and estimate a log stellar mass of $10.65\pm0.14$ and star formation rate up to $179\pm40$\msun\ yr$^{-1}$. We measure the total OH luminosity to $\log(L/L_{\odot}) = 3.90\pm0.03$, putting it on the cusp of the highest luminosity OH sources, OH gigamasers. The extreme star formation rate and infrared luminosity are consistent with the properties of the brightest known OHMs to date.  

The untargeted detection by AWES demonstrates the natural ability of wide-area \hi\ surveys with broad frequency bandwidth and complimentary redshift range to benefit spectral line searches for OH megamasers.
We predict that of order 110-150 sources will be found in the AWES survey, and speculate that of order 4-6 OH sources may be detected in satellite lines.  This increases the number of both known OHMs and OH satellite lines by more than 100\%.   The future is bright for OHMs: wide-area \hi\ surveys in addition to AWES, such as WALLABY expects to detect more than 100 OHMs over a different area of sky \citep{Koribalski20}.  In fact, Roberts, Darling, \& Baker (2021; submitted), suggest WALLABY may find even more OHMs, and based on our results, WALLABY will detect even more satellite line sources. Meanwhile, dedicated searches in absorption against background radio sources (Apertif SHARP; and MALS, \citealt{Gupta20}) will detect OH in different masing environments.

\begin{acknowledgements}

We are grateful to Tom Jarrrett for providing $k$-corrected WISE magnitudes and additional insight into the IR nature of the source; to James Allison for help in running the FLASHfinder tool to better understand the reliability of our measurements; and to James McBride for advice in recreating Figure 6b from his paper.  We also thank the anonymous referee for their comments which improved the quality of this paper.

HR and JD acknowledge support from the National Science Foundation grant AST-1814648.
EAKA is supported by the WISE research programme, which is financed by the Netherlands Organization for Scientific Research (NWO).
JMvdH acknowledges funding from the European Research Council under the European Union's Seventh Framework Programme (FP/2007-2013)/ERC Grant Agreement No. 291531 ('HIStoryNU').
JvL and LCO acknowledge funding from the European Research Council under the European Union's Seventh Framework Programme (FP/2007-2013) / ERC Grant Agreement n. 617199 (`ALERT'), and from Vici research programme `ARGO' with project number 639.043.815, financed by the Dutch Research Council (NWO).
DV acknowledges support from the Netherlands eScience Center (NLeSC) under grant ASDI.15.406.

      This work makes use of data from the Apertif system installed at the Westerbork Synthesis Radio Telescope owned by ASTRON. ASTRON, the Netherlands Institute for Radio Astronomy, is an institute of the Dutch Science Organisation (De Nederlandse Organisatie voor Wetenschappelijk Onderzoek, NWO). 
      
      This research has made use of the NASA/IPAC Extragalactic Database (NED), which is funded by the National Aeronautics and Space Administration and operated by the California Institute of Technology.
      
      The Pan-STARRS1 Surveys (PS1) and the PS1 public science archive have been made possible through contributions by the Institute for Astronomy, the University of Hawaii, the Pan-STARRS Project Office, the Max-Planck Society and its participating institutes, the Max Planck Institute for Astronomy, Heidelberg and the Max Planck Institute for Extraterrestrial Physics, Garching, The Johns Hopkins University, Durham University, the University of Edinburgh, the Queen's University Belfast, the Harvard-Smithsonian Center for Astrophysics, the Las Cumbres Observatory Global Telescope Network Incorporated, the National Central University of Taiwan, the Space Telescope Science Institute, the National Aeronautics and Space Administration under Grant No. NNX08AR22G issued through the Planetary Science Division of the NASA Science Mission Directorate, the National Science Foundation Grant No. AST-1238877, the University of Maryland, Eotvos Lorand University (ELTE), the Los Alamos National Laboratory, and the Gordon and Betty Moore Foundation. This research made use of Astropy,\footnote{http://www.astropy.org} a community-developed core Python package for Astronomy \citep{astropy2013, astropy18}.
\end{acknowledgements}

\bibliographystyle{aa}
\bibliography{ms_ohmaser_accepted}

\begin{thebibliography}{68}
\expandafter\ifx\csname natexlab\endcsname\relax\def\natexlab#1{#1}\fi

\bibitem[{{Adams} \& {van Leeuwen}(2019)}]{Adams19}
{Adams}, E. A.~K. \& {van Leeuwen}, J. 2019, Nature Astronomy, 3, 188

\bibitem[{{Allison} {et~al.}(2014){Allison}, {Sadler}, \& {Meekin}}]{Allison14}
{Allison}, J.~R., {Sadler}, E.~M., \& {Meekin}, A.~M. 2014, \mnras, 440, 696

\bibitem[{{Allison} {et~al.}(2012){Allison}, {Sadler}, \&
  {Whiting}}]{Allison12}
{Allison}, J.~R., {Sadler}, E.~M., \& {Whiting}, M.~T. 2012, \pasa, 29, 221

\bibitem[{{Astropy Collaboration} {et~al.}(2018){Astropy Collaboration},
  {Price-Whelan}, {Sip{\H{o}}cz}, {G{\"u}nther}, {Lim}, {Crawford}, {Conseil},
  {Shupe}, {Craig}, {Dencheva}, {Ginsburg}, {Vand erPlas}, {Bradley},
  {P{\'e}rez-Su{\'a}rez}, {de Val-Borro}, {Aldcroft}, {Cruz}, {Robitaille},
  {Tollerud}, {Ardelean}, {Babej}, {Bach}, {Bachetti}, {Bakanov}, {Bamford},
  {Barentsen}, {Barmby}, {Baumbach}, {Berry}, {Biscani}, {Boquien}, {Bostroem},
  {Bouma}, {Brammer}, {Bray}, {Breytenbach}, {Buddelmeijer}, {Burke},
  {Calderone}, {Cano Rodr{\'\i}guez}, {Cara}, {Cardoso}, {Cheedella}, {Copin},
  {Corrales}, {Crichton}, {D'Avella}, {Deil}, {Depagne}, {Dietrich}, {Donath},
  {Droettboom}, {Earl}, {Erben}, {Fabbro}, {Ferreira}, {Finethy}, {Fox},
  {Garrison}, {Gibbons}, {Goldstein}, {Gommers}, {Greco}, {Greenfield},
  {Groener}, {Grollier}, {Hagen}, {Hirst}, {Homeier}, {Horton}, {Hosseinzadeh},
  {Hu}, {Hunkeler}, {Ivezi{\'c}}, {Jain}, {Jenness}, {Kanarek}, {Kendrew},
  {Kern}, {Kerzendorf}, {Khvalko}, {King}, {Kirkby}, {Kulkarni}, {Kumar},
  {Lee}, {Lenz}, {Littlefair}, {Ma}, {Macleod}, {Mastropietro}, {McCully},
  {Montagnac}, {Morris}, {Mueller}, {Mumford}, {Muna}, {Murphy}, {Nelson},
  {Nguyen}, {Ninan}, {N{\"o}the}, {Ogaz}, {Oh}, {Parejko}, {Parley}, {Pascual},
  {Patil}, {Patil}, {Plunkett}, {Prochaska}, {Rastogi}, {Reddy Janga},
  {Sabater}, {Sakurikar}, {Seifert}, {Sherbert}, {Sherwood-Taylor}, {Shih},
  {Sick}, {Silbiger}, {Singanamalla}, {Singer}, {Sladen}, {Sooley},
  {Sornarajah}, {Streicher}, {Teuben}, {Thomas}, {Tremblay}, {Turner},
  {Terr{\'o}n}, {van Kerkwijk}, {de la Vega}, {Watkins}, {Weaver}, {Whitmore},
  {Woillez}, {Zabalza}, \& {Astropy Contributors}}]{astropy18}
{Astropy Collaboration}, {Price-Whelan}, A.~M., {Sip{\H{o}}cz}, B.~M., {et~al.}
  2018, \aj, 156, 123

\bibitem[{{Astropy Collaboration} {et~al.}(2013){Astropy Collaboration},
  {Robitaille}, {Tollerud}, {Greenfield}, {Droettboom}, {Bray}, {Aldcroft},
  {Davis}, {Ginsburg}, {Price-Whelan}, {Kerzendorf}, {Conley}, {Crighton},
  {Barbary}, {Muna}, {Ferguson}, {Grollier}, {Parikh}, {Nair}, {Unther},
  {Deil}, {Woillez}, {Conseil}, {Kramer}, {Turner}, {Singer}, {Fox}, {Weaver},
  {Zabalza}, {Edwards}, {Azalee Bostroem}, {Burke}, {Casey}, {Crawford},
  {Dencheva}, {Ely}, {Jenness}, {Labrie}, {Lim}, {Pierfederici}, {Pontzen},
  {Ptak}, {Refsdal}, {Servillat}, \& {Streicher}}]{astropy2013}
{Astropy Collaboration}, {Robitaille}, T.~P., {Tollerud}, E.~J., {et~al.} 2013,
  \aap, 558, A33

\bibitem[{{Baan}(1999)}]{Baan99}
{Baan}, W.~A. 1999, in IAU Symposium, Vol. 194, Activity in Galaxies and
  Related Phenomena, ed. Y.~{Terzian}, E.~{Khachikian}, \& D.~{Weedman}, 46

\bibitem[{{Baan} {et~al.}(1992{\natexlab{a}}){Baan}, {Haschick}, \&
  {Henkel}}]{Baan92a}
{Baan}, W.~A., {Haschick}, A., \& {Henkel}, C. 1992{\natexlab{a}}, \aj, 103,
  728

\bibitem[{{Baan} \& {Haschick}(1987)}]{Baan87}
{Baan}, W.~A. \& {Haschick}, A.~D. 1987, \apj, 318, 139

\bibitem[{{Baan} {et~al.}(1989){Baan}, {Haschick}, \& {Henkel}}]{Baan89}
{Baan}, W.~A., {Haschick}, A.~D., \& {Henkel}, C. 1989, \apj, 346, 680

\bibitem[{{Baan} \& {Kl{\"o}ckner}(2001)}]{Baan01}
{Baan}, W.~A. \& {Kl{\"o}ckner}, H.-R. 2001, in Astronomical Society of the
  Pacific Conference Series, Vol. 249, The Central Kiloparsec of Starbursts and
  AGN: The La Palma Connection, ed. J.~H. {Knapen}, J.~E. {Beckman},
  I.~{Shlosman}, \& T.~J. {Mahoney}, 639

\bibitem[{{Baan} {et~al.}(1992{\natexlab{b}}){Baan}, {Rhoads}, {Fisher},
  {Altschuler}, \& {Haschick}}]{Baan92b}
{Baan}, W.~A., {Rhoads}, J., {Fisher}, K., {Altschuler}, D.~R., \& {Haschick},
  A. 1992{\natexlab{b}}, \apjl, 396, L99

\bibitem[{{Baan} {et~al.}(1982){Baan}, {Wood}, \& {Haschick}}]{Baan82}
{Baan}, W.~A., {Wood}, P.~A.~D., \& {Haschick}, A.~D. 1982, \apjl, 260, L49

\bibitem[{{Briggs}(1998)}]{Briggs98}
{Briggs}, F.~H. 1998, \aap, 336, 815

\bibitem[{{Buat} {et~al.}(2007){Buat}, {Takeuchi}, {Iglesias-P{\'a}ramo}, {Xu},
  {Burgarella}, {Boselli}, {Barlow}, {Bianchi}, {Donas}, {Forster}, {Friedman},
  {Heckman}, {Lee}, {Madore}, {Martin}, {Milliard}, {Morissey}, {Neff}, {Rich},
  {Schiminovich}, {Seibert}, {Small}, {Szalay}, {Welsh}, {Wyder}, \&
  {Yi}}]{Buat07}
{Buat}, V., {Takeuchi}, T.~T., {Iglesias-P{\'a}ramo}, J., {et~al.} 2007, \apjs,
  173, 404

\bibitem[{{Chambers} {et~al.}(2016){Chambers}, {Magnier}, {Metcalfe},
  {Flewelling}, {Huber}, {Waters}, {Denneau}, {Draper}, {Farrow}, {Finkbeiner},
  {Holmberg}, {Koppenhoefer}, {Price}, {Rest}, {Saglia}, {Schlafly}, {Smartt},
  {Sweeney}, {Wainscoat}, {Burgett}, {Chastel}, {Grav}, {Heasley}, {Hodapp},
  {Jedicke}, {Kaiser}, {Kudritzki}, {Luppino}, {Lupton}, {Monet}, {Morgan},
  {Onaka}, {Shiao}, {Stubbs}, {Tonry}, {White}, {Ba{\~n}ados}, {Bell},
  {Bender}, {Bernard}, {Boegner}, {Boffi}, {Botticella}, {Calamida},
  {Casertano}, {Chen}, {Chen}, {Cole}, {Deacon}, {Frenk}, {Fitzsimmons},
  {Gezari}, {Gibbs}, {Goessl}, {Goggia}, {Gourgue}, {Goldman}, {Grant},
  {Grebel}, {Hambly}, {Hasinger}, {Heavens}, {Heckman}, {Henderson}, {Henning},
  {Holman}, {Hopp}, {Ip}, {Isani}, {Jackson}, {Keyes}, {Koekemoer}, {Kotak},
  {Le}, {Liska}, {Long}, {Lucey}, {Liu}, {Martin}, {Masci}, {McLean}, {Mindel},
  {Misra}, {Morganson}, {Murphy}, {Obaika}, {Narayan}, {Nieto-Santisteban},
  {Norberg}, {Peacock}, {Pier}, {Postman}, {Primak}, {Rae}, {Rai}, {Riess},
  {Riffeser}, {Rix}, {R{\"o}ser}, {Russel}, {Rutz}, {Schilbach}, {Schultz},
  {Scolnic}, {Strolger}, {Szalay}, {Seitz}, {Small}, {Smith}, {Soderblom},
  {Taylor}, {Thomson}, {Taylor}, {Thakar}, {Thiel}, {Thilker}, {Unger},
  {Urata}, {Valenti}, {Wagner}, {Walder}, {Walter}, {Watters}, {Werner},
  {Wood-Vasey}, \& {Wyse}}]{Chambers2016}
{Chambers}, K.~C., {Magnier}, E.~A., {Metcalfe}, N., {et~al.} 2016, arXiv
  e-prints, arXiv:1612.05560

\bibitem[{{Chengalur} \& {Kanekar}(2003)}]{Chengalur03}
{Chengalur}, J.~N. \& {Kanekar}, N. 2003, \prl, 91, 241302

\bibitem[{{Clements} {et~al.}(1996){Clements}, {Sutherland}, {McMahon}, \&
  {Saunders}}]{Clements96}
{Clements}, D.~L., {Sutherland}, W.~J., {McMahon}, R.~G., \& {Saunders}, W.
  1996, \mnras, 279, 477

\bibitem[{{Cluver} {et~al.}(2017){Cluver}, {Jarrett}, {Dale}, {Smith},
  {August}, \& {Brown}}]{Cluver17}
{Cluver}, M.~E., {Jarrett}, T.~H., {Dale}, D.~A., {et~al.} 2017, \apj, 850, 68

\bibitem[{{Cluver} {et~al.}(2014){Cluver}, {Jarrett}, {Hopkins}, {Driver},
  {Liske}, {Gunawardhana}, {Taylor}, {Robotham}, {Alpaslan}, {Baldry}, {Brown},
  {Peacock}, {Popescu}, {Tuffs}, {Bauer}, {Bland -Hawthorn}, {Colless},
  {Holwerda}, {Lara-L{\'o}pez}, {Leschinski}, {L{\'o}pez-S{\'a}nchez},
  {Norberg}, {Owers}, {Wang}, \& {Wilkins}}]{Cluver14}
{Cluver}, M.~E., {Jarrett}, T.~H., {Hopkins}, A.~M., {et~al.} 2014, \apj, 782,
  90

\bibitem[{{Darling}(2003)}]{Darling03}
{Darling}, J. 2003, \prl, 91, 011301

\bibitem[{{Darling}(2004)}]{Darling04}
{Darling}, J. 2004, \apj, 612, 58

\bibitem[{{Darling}(2007)}]{Darling07}
{Darling}, J. 2007, \apjl, 669, L9

\bibitem[{{Darling} \& {Giovanelli}(2001)}]{Darling01}
{Darling}, J. \& {Giovanelli}, R. 2001, \aj, 121, 1278

\bibitem[{{Darling} \& {Giovanelli}(2002)}]{Darling02}
{Darling}, J. \& {Giovanelli}, R. 2002, \aj, 124, 100

\bibitem[{{Farrah} {et~al.}(2001){Farrah}, {Rowan-Robinson}, {Oliver},
  {Serjeant}, {Borne}, {Lawrence}, {Lucas}, {Bushouse}, \& {Colina}}]{Farrah01}
{Farrah}, D., {Rowan-Robinson}, M., {Oliver}, S., {et~al.} 2001, \mnras, 326,
  1333

\bibitem[{{Fernandez} {et~al.}(2010){Fernandez}, {Momjian}, {Salter}, \&
  {Ghosh}}]{Fernandez10}
{Fernandez}, M.~X., {Momjian}, E., {Salter}, C.~J., \& {Ghosh}, T. 2010, \aj,
  139, 2066

\bibitem[{{Flambaum} \& {Kozlov}(2007)}]{Flambaum07}
{Flambaum}, V.~V. \& {Kozlov}, M.~G. 2007, \prl, 98, 240801

\bibitem[{{Frayer} {et~al.}(1998){Frayer}, {Seaquist}, \& {Frail}}]{Frayer98}
{Frayer}, D.~T., {Seaquist}, E.~R., \& {Frail}, D.~A. 1998, \aj, 115, 559

\bibitem[{{Gupta} {et~al.}(2020){Gupta}, {Jagannathan}, {Srianand},
  {Bhatnagar}, {Noterdaeme}, {Combes}, {Petitjean}, {Jose}, {Pandey}, {Kaski},
  {Baker}, {Balashev}, {Boettcher}, {Chen}, {Cress}, {Dutta}, {Goedhart},
  {Heald}, {J{\'o}zsa}, {Kamau}, {Kamphuis}, {Kerp}, {Kl{\"o}ckner}, {Knowles},
  {Krishnan}, {Krogager}, {Kulkarni}, {Momjian}, {Moodley}, {Passmoor},
  {Schr{\"o}eder}, {Sekhar}, {Sikhosana}, {Wagenveld}, \& {Wong}}]{Gupta20}
{Gupta}, N., {Jagannathan}, P., {Srianand}, R., {et~al.} 2020, arXiv e-prints,
  arXiv:2007.04347

\bibitem[{{Hao} {et~al.}(2011){Hao}, {Kennicutt}, {Johnson}, {Calzetti},
  {Dale}, \& {Moustakas}}]{Hao11}
{Hao}, C.-N., {Kennicutt}, R.~C., {Johnson}, B.~D., {et~al.} 2011, \apj, 741,
  124

\bibitem[{{Haynes} {et~al.}(2018){Haynes}, {Giovanelli}, {Kent}, {Adams},
  {Balonek}, {Craig}, {Fertig}, {Finn}, {Giovanardi}, {Hallenbeck}, {Hess},
  {Hoffman}, {Huang}, {Jones}, {Koopmann}, {Kornreich}, {Leisman}, {Miller},
  {Moorman}, {O'Connor}, {O'Donoghue}, {Papastergis}, {Troischt}, {Stark}, \&
  {Xiao}}]{Haynes18}
{Haynes}, M.~P., {Giovanelli}, R., {Kent}, B.~R., {et~al.} 2018, \apj, 861, 49

\bibitem[{{Henkel} {et~al.}(1991){Henkel}, {Baan}, \&
  {Mauersberger}}]{Henkel91}
{Henkel}, C., {Baan}, W.~A., \& {Mauersberger}, R. 1991, \aapr, 3, 47

\bibitem[{{Houck} {et~al.}(1985){Houck}, {Schneider}, {Danielson}, {Beichman},
  {Lonsdale}, {Neugebauer}, \& {Soifer}}]{Houck85}
{Houck}, J.~R., {Schneider}, D.~P., {Danielson}, G.~E., {et~al.} 1985, \apjl,
  290, L5

\bibitem[{{Jansen} {et~al.}(2011){Jansen}, {Xu}, {Kleiner}, {Ubachs}, \&
  {Bethlem}}]{Jansen11}
{Jansen}, P., {Xu}, L.-H., {Kleiner}, I., {Ubachs}, W., \& {Bethlem}, H.~L.
  2011, \prl, 106, 100801

\bibitem[{{Jarrett} {et~al.}(2019){Jarrett}, {Cluver}, {Brown}, {Dale}, {Tsai},
  \& {Masci}}]{Jarrett19}
{Jarrett}, T.~H., {Cluver}, M.~E., {Brown}, M.~J.~I., {et~al.} 2019, \apjs,
  245, 25

\bibitem[{{Kanekar} {et~al.}(2005){Kanekar}, {Carilli}, {Langston}, {Rocha},
  {Combes}, {Subrahmanyan}, {Stocke}, {Menten}, {Briggs}, \&
  {Wiklind}}]{Kanekar05}
{Kanekar}, N., {Carilli}, C.~L., {Langston}, G.~I., {et~al.} 2005, \prl, 95,
  261301

\bibitem[{{Kanekar} {et~al.}(2018){Kanekar}, {Ghosh}, \&
  {Chengalur}}]{Kanekar18}
{Kanekar}, N., {Ghosh}, T., \& {Chengalur}, J.~N. 2018, \prl, 120, 061302

\bibitem[{{Kanekar} {et~al.}(2012){Kanekar}, {Langston}, {Stocke}, {Carilli},
  \& {Menten}}]{Kanekar12}
{Kanekar}, N., {Langston}, G.~I., {Stocke}, J.~T., {Carilli}, C.~L., \&
  {Menten}, K.~M. 2012, \apjl, 746, L16

\bibitem[{{Kennicutt} \& {Evans}(2012)}]{Kennicutt12}
{Kennicutt}, R.~C. \& {Evans}, N.~J. 2012, \araa, 50, 531

\bibitem[{{Koribalski} {et~al.}(2020){Koribalski}, {Staveley-Smith},
  {Westmeier}, {Serra}, {Spekkens}, {Wong}, {Lee-Waddell}, {Lagos},
  {Obreschkow}, {Ryan-Weber}, {Zwaan}, {Kilborn}, {Bekiaris}, {Bekki},
  {Bigiel}, {Boselli}, {Bosma}, {Catinella}, {Chauhan}, {Cluver}, {Colless},
  {Courtois}, {Crain}, {de Blok}, {D{\'e}nes}, {Duffy}, {Elagali}, {Fluke},
  {For}, {Heald}, {Henning}, {Hess}, {Holwerda}, {Howlett}, {Jarrett}, {Jones},
  {Jones}, {J{\'o}zsa}, {Jurek}, {J{\"u}tte}, {Kamphuis}, {Karachentsev},
  {Kerp}, {Kleiner}, {Kraan-Korteweg}, {L{\'o}pez-S{\'a}nchez}, {Madrid},
  {Meyer}, {Mould}, {Murugeshan}, {Norris}, {Oh}, {Oosterloo}, {Popping},
  {Putman}, {Reynolds}, {Rhee}, {Robotham}, {Ryder}, {Schr{\"o}der}, {Shao},
  {Stevens}, {Taylor}, {van{\^A} der Hulst}, {Verdes-Montenegro}, {Wakker},
  {Wang}, {Whiting}, {Winkel}, \& {Wolf}}]{Koribalski20}
{Koribalski}, B.~S., {Staveley-Smith}, L., {Westmeier}, T., {et~al.} 2020,
  \apss, 365, 118

\bibitem[{{Lo}(2005)}]{Lo05}
{Lo}, K.~Y. 2005, \araa, 43, 625

\bibitem[{{Lockett} \& {Elitzur}(2008)}]{Lockett08}
{Lockett}, P. \& {Elitzur}, M. 2008, \apj, 677, 985

\bibitem[{{Lonsdale}(2002)}]{Lonsdale02}
{Lonsdale}, C.~J. 2002, in IAU Symposium, Vol. 206, Cosmic Masers: From
  Proto-Stars to Black Holes, ed. V.~{Migenes} \& M.~J. {Reid}, 413

\bibitem[{Lotz {et~al.}(2011)Lotz, Jonsson, Cox, Croton, Primack, Somerville,
  \& Stewart}]{Lotz2011}
Lotz, J.~M., Jonsson, P., Cox, T.~J., {et~al.} 2011, Astrophysical Journal,
  742, 1

\bibitem[{{McBride} {et~al.}(2013){McBride}, {Heiles}, \&
  {Elitzur}}]{McBride13}
{McBride}, J., {Heiles}, C., \& {Elitzur}, M. 2013, \apj, 774, 35

\bibitem[{{McKean} \& {Roy}(2009)}]{McKean09}
{McKean}, J. \& {Roy}, A.~L. 2009, in Panoramic Radio Astronomy: Wide-field 1-2
  GHz Research on Galaxy Evolution, 60

\bibitem[{{Moshir} \& {et al.}(1990)}]{Moshir90}
{Moshir}, M. \& {et al.} 1990, IRAS Faint Source Catalogue, 0

\bibitem[{{Murphy} {et~al.}(2011){Murphy}, {Condon}, {Schinnerer}, {Kennicutt},
  {Calzetti}, {Armus}, {Helou}, {Turner}, {Aniano}, {Beir{\~a}o}, {Bolatto},
  {Brandl}, {Croxall}, {Dale}, {Donovan Meyer}, {Draine}, {Engelbracht},
  {Hunt}, {Hao}, {Koda}, {Roussel}, {Skibba}, \& {Smith}}]{Murphy11}
{Murphy}, E.~J., {Condon}, J.~J., {Schinnerer}, E., {et~al.} 2011, \apj, 737,
  67

\bibitem[{{Norris} {et~al.}(1989){Norris}, {Gardner}, {Whiteoak}, {Allen}, \&
  {Roche}}]{Norris89}
{Norris}, R.~P., {Gardner}, F.~F., {Whiteoak}, J.~B., {Allen}, D.~A., \&
  {Roche}, P.~F. 1989, \mnras, 237, 673

\bibitem[{{Parra} {et~al.}(2005){Parra}, {Conway}, {Elitzur}, \&
  {Pihlstr{\"o}m}}]{Parra05}
{Parra}, R., {Conway}, J.~E., {Elitzur}, M., \& {Pihlstr{\"o}m}, Y.~M. 2005,
  \aap, 443, 383

\bibitem[{{Radford}(1964)}]{Radford64}
{Radford}, H.~E. 1964, \prl, 13, 534

\bibitem[{{Robinson} \& {McGee}(1967)}]{Robinson67}
{Robinson}, B.~J. \& {McGee}, R.~X. 1967, \araa, 5, 183

\bibitem[{{Sanders} \& {Mirabel}(1996)}]{Sanders96}
{Sanders}, D.~B. \& {Mirabel}, I.~F. 1996, \araa, 34, 749

\bibitem[{{Sault} {et~al.}(2011){Sault}, {Teuben}, \& {Wright}}]{Sault11}
{Sault}, R.~J., {Teuben}, P., \& {Wright}, M. C.~H. 2011, {MIRIAD:
  Multi-channel Image Reconstruction, Image Analysis, and Display}

\bibitem[{{Sault} {et~al.}(1995){Sault}, {Teuben}, \& {Wright}}]{Sault95}
{Sault}, R.~J., {Teuben}, P.~J., \& {Wright}, M.~C.~H. 1995, in Astronomical
  Society of the Pacific Conference Series, Vol.~77, Astronomical Data Analysis
  Software and Systems IV, ed. R.~A. {Shaw}, H.~E. {Payne}, \& J.~J.~E.
  {Hayes}, 433

\bibitem[{{Schulz} {et~al.}(2020){Schulz}, {Dijkema}, \& {Molenaar}}]{Schulz20}
{Schulz}, R., {Dijkema}, T.~J., \& {Molenaar}, G. 2020, {Apercal: Pipeline for
  the Westerbork Synthesis Radio Telescope Apertif upgrade}

\bibitem[{{Seaquist} {et~al.}(1997){Seaquist}, {Frayer}, \&
  {Frail}}]{Seaquist97}
{Seaquist}, E.~R., {Frayer}, D.~T., \& {Frail}, D.~A. 1997, \apjl, 487, L131

\bibitem[{{Serra} {et~al.}(2015){Serra}, {Westmeier}, {Giese}, {Jurek},
  {Fl{\"o}er}, {Popping}, {Winkel}, {van der Hulst}, {Meyer}, {Koribalski},
  {Staveley-Smith}, \& {Courtois}}]{Serra15}
{Serra}, P., {Westmeier}, T., {Giese}, N., {et~al.} 2015, \mnras, 448, 1922

\bibitem[{{Skinner} {et~al.}(1997){Skinner}, {Smith}, {Sturm}, {Barlow},
  {Cohen}, \& {Stacey}}]{Skinner97}
{Skinner}, C.~J., {Smith}, H.~A., {Sturm}, E., {et~al.} 1997, \nat, 386, 472

\bibitem[{{Staveley-Smith} {et~al.}(1989){Staveley-Smith}, {Allen}, {Chapman},
  {Norris}, \& {Whiteoak}}]{StaveleySmith89}
{Staveley-Smith}, L., {Allen}, D.~A., {Chapman}, J.~M., {Norris}, R.~P., \&
  {Whiteoak}, J.~B. 1989, \nat, 337, 625

\bibitem[{{Staveley-Smith} {et~al.}(1992){Staveley-Smith}, {Norris}, {Chapman},
  {Allen}, {Whiteoak}, \& {Roy}}]{StaveleySmith92}
{Staveley-Smith}, L., {Norris}, R.~P., {Chapman}, J.~M., {et~al.} 1992, \mnras,
  258, 725

\bibitem[{{Suess} {et~al.}(2016){Suess}, {Darling}, {Haynes}, \&
  {Giovanelli}}]{Suess2016}
{Suess}, K.~A., {Darling}, J., {Haynes}, M.~P., \& {Giovanelli}, R. 2016,
  \mnras, 459, 220

\bibitem[{{Varshalovich} \& {Potekhin}(1995)}]{Varshalovich95}
{Varshalovich}, D.~A. \& {Potekhin}, A.~Y. 1995, \ssr, 74, 259

\bibitem[{Virtanen {et~al.}(2020)Virtanen, Gommers, Oliphant, Haberland, Reddy,
  Cournapeau, Burovski, Peterson, Weckesser, Bright, {van der Walt}, Brett,
  Wilson, Millman, Mayorov, Nelson, Jones, Kern, Larson, Carey, Polat, Feng,
  Moore, {VanderPlas}, Laxalde, Perktold, Cimrman, Henriksen, Quintero, Harris,
  Archibald, Ribeiro, Pedregosa, {van Mulbregt}, \& {SciPy 1.0
  Contributors}}]{Virtanen20}
Virtanen, P., Gommers, R., Oliphant, T.~E., {et~al.} 2020, Nature Methods, 17,
  261

\bibitem[{{Webb} {et~al.}(2001){Webb}, {Murphy}, {Flambaum}, {Dzuba}, {Barrow},
  {Churchill}, {Prochaska}, \& {Wolfe}}]{Webb01}
{Webb}, J.~K., {Murphy}, M.~T., {Flambaum}, V.~V., {et~al.} 2001, \prl, 87,
  091301

\bibitem[{{Willett}(2012)}]{Willett12}
{Willett}, K.~W. 2012, in IAU Symposium, Vol. 287, Cosmic Masers - from OH to
  H0, ed. R.~S. {Booth}, W.~H.~T. {Vlemmings}, \& E.~M.~L. {Humphreys},
  345--349

\bibitem[{{Willett} {et~al.}(2011){Willett}, {Darling}, {Spoon},
  {Charmandaris}, \& {Armus}}]{Willett11}
{Willett}, K.~W., {Darling}, J., {Spoon}, H. W.~W., {Charmandaris}, V., \&
  {Armus}, L. 2011, \apj, 730, 56

\bibitem[{{Zhang} {et~al.}(2014){Zhang}, {Wang}, {Di}, {Zhu}, {Guo}, \&
  {Wang}}]{Zhang14}
{Zhang}, J.~S., {Wang}, J.~Z., {Di}, G.~X., {et~al.} 2014, \aap, 570, A110

\end{thebibliography}

\begin{appendix} 

\section{Reliability of the 1612 MHz line}
\label{sect:app}

\subsection{Matched filter}

\begin{figure*}
    \centering
    \includegraphics[width=\textwidth]{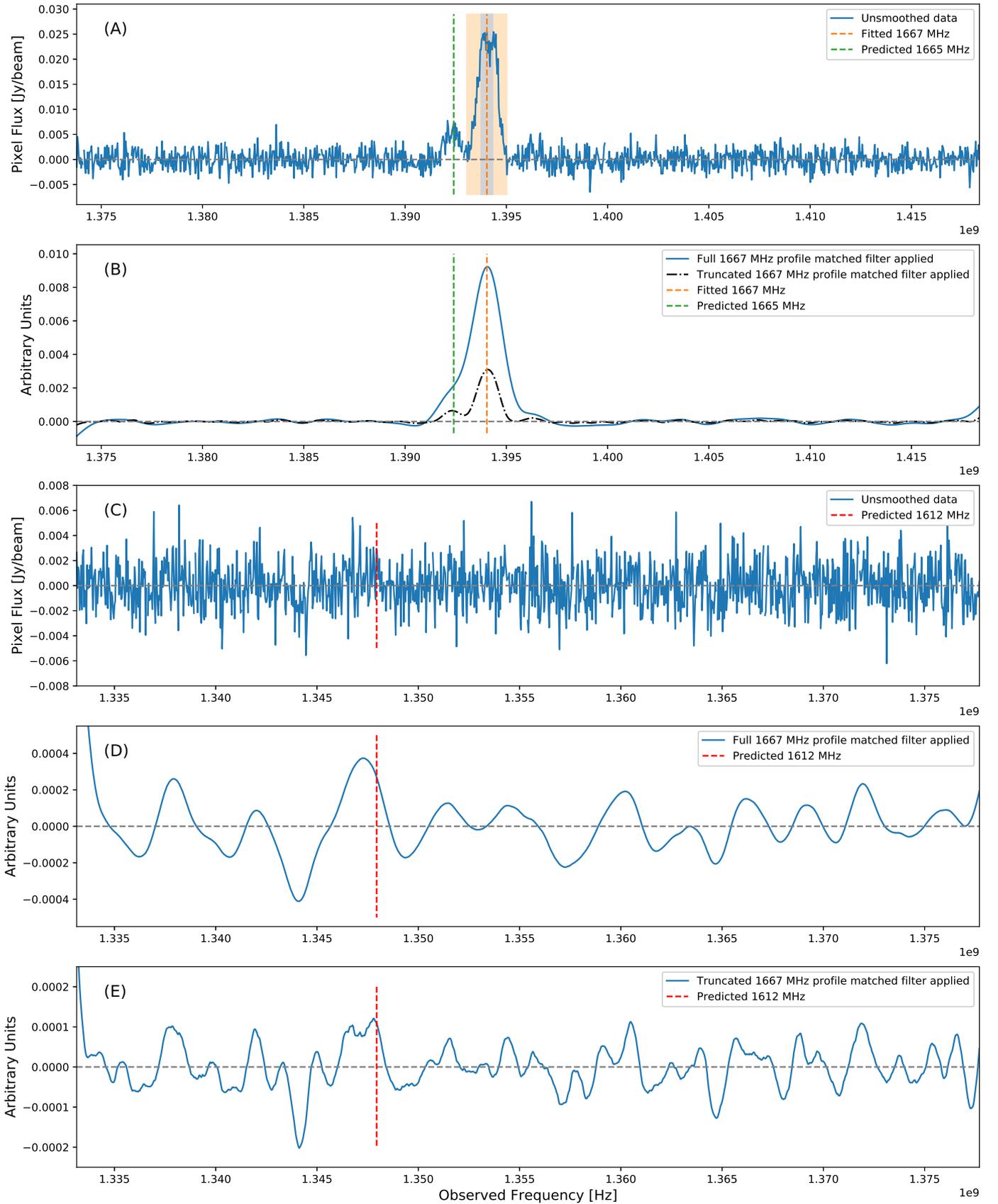}
    \caption{Panel A: Full resolution (unsmoothed) pixel spectrum showing the 1665 and 1667 MHz lines. Orange shaded region shows the extent of the OH 1667 MHz profile for the full matched filter. Gray shaded region shows extent of the OH 1667 MHz profile that has been used for the truncated filter.  It matches the width over which the 1612 MHz line was measured in Section \ref{sect:results}.  Panel B: The matched filters applied to the unsmoothed data. Panel C: Full resolution pixel spectrum covering the predicted frequency of the 1612 MHz line.  Panel D: The full 1667 profile matched filter applied to the unsmoothed 1612 MHz data.  Panel E: The truncated 1667 MHz profile matched filter applied to the unsmoothed 1612 MHz data.}
    \label{fig:matchedfilter}
\end{figure*}

In order to test the reliability of the 1612 MHz detection we performed a test using matched filtering.  Here we present the matched filtering results on the full spectral resolution (unsmoothed) data, however we have also performed the exact same tests on the smoothed data presented in this work and found the resulting plots are essentially identical.

The shape of the 1612 MHz line is not known \textit{a priori}, however we start with the assumption that the line shapes are approximately the same for all the OH maser lines.  This implies that all lines trace the same gas components, although in practice that may not be the case. We take the 1667 MHz profile over the channel range indicated in Figure \ref{fig:spectra}, and reproduced in Figure \ref{fig:matchedfilter}, as an initial guess for the matched filter and apply it using \texttt{scipy.signal.filtfilt} \citep{Virtanen20}.  To check the implementation, we first apply the matched filter to the 1667 and 1665 MHz line spectrum.  The signal-to-noise is maximized exactly where the 1667 MHz line is detected, as expected, although we note that the 1665 MHz signal is strongly blended with the 1667 MHz signal (blue line in Panel B of Figure \ref{fig:matchedfilter}).

We attempted to improve on this by truncating the matched filter at the estimated width of the 1612 MHz line (the line center $\pm8$ channels corresponding to the gray shaded region of Figure \ref{fig:matchedfilter}).  When we apply the truncated matched filter to the 1665 and 1667 MHz spectrum (black line in Figure \ref{fig:matchedfilter}), the 1665 MHz signal is now distinct and its peak is also very well localized with the predicted frequency of the 1665 MHz line.  

The Panel D of Figure \ref{fig:matchedfilter} shows the result of cross correlating the full matched filter with the 1612 MHz spectrum.  The most significant feature, positive or negative, is associated with where we expect the redshift 1612 MHz line to fall.  (The rise at lowest frequencies is an edge effect.)  In fact, using the entire 1667 MHz profile for the matched filter may be too broad: 
while this filter maximizes the SNR, its localization is poor. 

Panel E of Figure \ref{fig:matchedfilter} shows the cross correlation of 1612 MHz spectrum with this truncated matched filter.  The strongest overall feature in the matched filter correlation is negative, and its profile shape appears noise-like.  The strongest positive feature is still associated with where we expect the 1612 MHz line to fall, but now the localization is significantly better, and remarkably well aligned with the predicted position of the 1612 MHz line. The shape of this positive feature also appears qualitatively different than the noise features around it.

Finally, we repeated the work described above with a Gaussian kernel fit to the 1667 MHz profile and a truncated Gaussian profile, which is simply the peak of the same Gaussian, now restricted to the inner $\pm8$ channels.  We find that the results for the full Gaussian are exactly the same as for the full matched filter.  For the truncated Gaussian, the result is to smooth out the very small (channel by channel) variations seen in Panel E but it does not change the large scale variations seen in the plot, or our results.

What we conclude from the six combinations of applying a full or truncated matched filter, or a Gausssian filter to both the full resolution and the smoothed pixel spectra are that the exact shape of the filter does not matter in terms of identifying the most significant features in the spectra.  However, the width of the filter is important in localizing these features (e.g.~Panel B).  Panel (E) is suggestive that there is emission detected from the 1612 MHz line, but the conclusions are tentative and the measurement is still dominated by noise.

\subsection{FLASHfinder}

We performed a second test using the Bayesian line fitting tool FLASHfinder\footnote{https://github.com/drjamesallison/flash\_finder} \citep{Allison12,Allison14} with a strong redshift constraint.  
As with the matched filter, the tool was tested on the 1665 and 1667 MHz spectrum, in this case either assuming two Gaussian components, or assuming three Gaussian components with redshift priors of $z_{\odot}=0.19612\pm0.01$.  The three component Gaussian was overwhelmingly favoured with a difference in log-probability of $17.61\pm0.07$, in order to fit the double peak in the 1667MHz line.

FLASHfinder was then run on the 1612 MHz spectrum with an even stronger redshift prior of $z_{\odot}=0.19612\pm0.001$. This corresponds to an error of $\sim300$\kms, although the error in the redshift uncertainty measured by SoFiA for the 1667 MHz line is only 1\kms.  Unfortunately, the Bayesian tool disfavors a Gaussian component at the predicted location of the 1612 MHz line with a log probability of -0.27.  More optimistically phrased, the log probability estimates that there is a roughly 1 in 3 chance that we have detected the spectral line.

As a result of both the matched filter and the Bayesian analysis, we take the conservative approach throughout the paper of calling the 1612 MHz line measurement an upper limit rather than a tentative detection.

\end{appendix}

\end{document}